\documentclass[iop]{emulateapj}
\usepackage{multirow}
\usepackage{natbib}

\usepackage[latin1]{inputenc}
\usepackage{amsmath}
\usepackage{color}

\shorttitle{Calibrated UFig Simulations for DES}
\shortauthors{Bruderer et al.}

\begin{document}

\title{Calibrated Ultra Fast Image Simulations for the Dark Energy Survey}

\author{Claudio Bruderer\altaffilmark{1$\dagger$}}
\email{$\dagger$ claudio.bruderer@phys.ethz.ch}
\author{Chihway Chang\altaffilmark{1}}
\author{Alexandre Refregier\altaffilmark{1}}
\author{Adam Amara\altaffilmark{1}}
\author{Joel Berg\'{e}\altaffilmark{1,2}}
\author{Lukas Gamper\altaffilmark{1}}
\affil{\altaffilmark{1} Institute for Astronomy, Department of Physics, ETH Zurich, Wolfgang-Pauli-Strasse 27, 8093 Z\"urich, Switzerland}
\affil{\altaffilmark{2} ONERA - The French Aerospace Lab, 29 avenue de la Division Leclerc, 92320 Ch\^atillon, France}

\begin{abstract}
Weak lensing by large-scale structure is a powerful technique to probe the dark components of the universe. To understand the measurement process of weak lensing and the associated systematic effects, image simulations are becoming increasingly important. For this purpose we present a first implementation of the \textit{Monte Carlo Control Loops} \citep[\textit{MCCL}; ][]{Refregier:2014aa}, a coherent framework for studying systematic effects in weak lensing. It allows us to model and calibrate the shear measurement process using image simulations from the Ultra Fast Image Generator \citep[\textsc{UFig}; ][]{Berge:2013aa}. We apply this framework to a subset of the data taken during the Science Verification period (SV) of the Dark Energy Survey (DES). We calibrate the \textsc{UFig} simulations to be statistically consistent with DES images. We then perform tolerance analyses by perturbing the simulation parameters and study their impact on the shear measurement at the one-point level. This allows us to determine the relative importance of different input parameters to the simulations. For spatially constant systematic errors and six simulation parameters, the calibration of the simulation reaches the weak lensing precision needed for the DES SV survey area. Furthermore, we find a sensitivity of the shear measurement to the intrinsic ellipticity distribution, and an interplay between the magnitude-size and the pixel value diagnostics in constraining the noise model. This work is the first application of the \textit{MCCL} framework to data and shows how it can be used to methodically study the impact of systematics on the cosmic shear measurement.
\end{abstract}


\section{Introduction}\label{sec:introduction}
Within the last decades our picture of the Universe changed dramatically with the discovery of its accelerating expansion attributed to a mysterious dark energy. Together with dark matter they make up the dark sector of the Universe. The introduction of dark energy has led to the establishment of the $\Lambda$CDM-model as the current cosmological standard model. The model agrees well with observations from different cosmological probes \citep[e.g][]{Hinshaw:2013aa,Planck-Collaboration:2014aa}. Nonetheless, understanding the nature of the dark sector is one of cosmology's most pressing challenges.

Weak gravitational lensing \citep[for reviews see][]{Refregier:2003aa,Hoekstra:2008aa}, a distortion effect of galaxy shapes due to interloping structures along the line-of-sight, has a large potential to shed light onto the mystery of the dark sector \citep{Albrecht:2006aa}. It is a purely gravitational effect, and thus reacts in the same way to dark and baryonic matter. However, the induced distortions on galaxy shapes are very weak ($\sim$1\%). Therefore, galaxy shapes need to be measured to a very high precision for weak lensing to unleash its full potential as a cosmological probe \citep[e.g.][henceforth AR08]{Huterer:2006aa,Amara:2008aa}.

Many shape-measurement algorithms have been developed over the past two decades \citep[for an overview see][and references therein]{Zuntz:2013aa}. Image simulation plays an important role for calibrating and validating those methods. To test the performance of various shear measurement codes on simulated images, public challenges like the Shear TEsting Programs (STEP) \citep{Heymans:2006aa,Massey:2007aa} and the GRavitation lEnsing Accuracy Testing (GREAT) \citep{Bridle:2009aa,Kitching:2012aa,Mandelbaum:2014aa} were established. Valuable insight in the measurement process could be gained and significant progress was made. However, these challenges reaffirmed that a careful and rigorous treatment of systematic errors is essential to weak lensing as a cosmological probe.

Several large wide-field imaging surveys will come online in the next few years, including the Dark Energy Survey\footnote{http://www.darkenergysurvey.org/} (DES), the Kilo Degree Survey\footnote{http://kids.strw.leidenuniv.nl/} (KiDS), Hyper Suprime-Cam\footnote{http://www.naoj.org/Projects/HSC/index.html} (HSC), Euclid\footnote{http://sci.esa.int/euclid/}, and the Large Synoptic Survey Telescope\footnote{http://www.lsst.org/lsst/} (LSST). They will map out a large fraction of the Sky, yielding a wealth of data. In this work, we will focus on data from DES.

In this paper, we present an initial implementation of a novel shear measurement approach, the \textit{Monte Carlo Control Loops} \citep[\textit{MCCL}; ][henceforth RA13]{Refregier:2014aa} at the one-point statistics level. We use image simulations that pass through the same lensing measurement pipeline as the data to forward model the measurement process. In this approach, not only can the shear measurement be calibrated, the nature of the pipeline allows us to test the robustness of the calibration. The \textit{MCCL}-framework dynamically modifies the lensing pipeline and aims to provide a shear measurement with systematic errors smaller than the statistical errors for the survey being considered. For this purpose, we establish a set of three iterative \textit{Control Loops (CL)} which build upon each other. First, the simulations are tuned to be statistically consistent with data. Second, the lensing measurement is calibrated. Third, the robustness of the calibration is tested. We achieve this by perturbing the simulation parameters and recalibrating the measurement while keeping data and simulations statistically consistent. The uncertainty in the calibration due to the perturbed parameters gives the systematic error of the measurement.

This paper is organized as follows. In Section \ref{sec:shearMsrtProb}, we explain the main concept and requirements in using the \textit{MCCL} framework to tackle the shear measurement problem. In Section \ref{sec:data} we give a description of the DES SV data. The main features of \textsc{UFig} are described in Section \ref{sec:UFig}. We focus especially on the properties of the simulated galaxies, PSF, noise, and the shear field. In Section \ref{sec:method} we present the \textit{MCCL} framework and its implementation. We show in Section \ref{sec:results} different diagnostics of our calibrated image simulations for DES. Furthermore, the results of our first tentative analysis of the robustness of the shear measurement calibration are presented. We conclude in Section \ref{sec:conclusion}.

\section{\textit{MCCL} and the Shear Measurement Problem}\label{sec:shearMsrtProb}
The main goal of this paper is to tackle the weak lensing shear measurement problem using the \textit{MCCL} approach proposed by RA13. In this section we elaborate on the main concepts behind the \textit{MCCL} framework and how that translates into the specific implementations (see Section~\ref{sec:method}) carried out in this paper.

The first key concept is that all the \textit{CL}s need to be specifically ``controlled'' by certain criteria, or targets. For example, in our first \textit{CL}, we define criteria within which we view the simulations and the data to be statistically consistent. When the simulations satisfy the requirements, we leave the loop and continue on to the next step. The overall target that controls the entire framework is naturally tied to the science goals of the framework. In our case, the target is producing shear measurements that are accurate within statistical errors of the DES dataset of interest.

We choose to set the main target of this paper using results from AR08. First, we parametrize our measured shear to be linearly related to the true underlying shear via  
\begin{equation}\label{eq:defMultAddBias}
    \gamma^{obs}_{i} = (1+m_{i})\gamma^{t}_{i}+c_{i} + N,
\end{equation}
where $\gamma^{t}_{i}$ is the true shear, $\gamma^{obs}_{i}$ the estimated shear. $m_{i}$ and $c_{i}$ are the multiplicative and additive biases, and N is a noise term that averages out for large numbers. Then, according to AR08, in order for the shear measurement not to be systematics-dominated, we would require $m_{i}\lesssim0.025$ and $c_{i}\lesssim1.65\cdot10^{-3}$ for a DES SV-like 200 deg$^{2}$ survey, and $m_{i}\lesssim0.005$ and $c_{i}\lesssim0.75\cdot10^{-3}$ for the full 5000 deg$^{2}$ DES survey. While these upper limits were derived by AR08 for two-point statistics, they place requirements on one-point statistics, namely that the absolute means of $m_{i}$ and $c_{i}$ must stay below the limits stated. This can be thought of as the requirements for the case of spatially constant systematic effects.

Following the logic above, the second key concept is that the \textit{MCCL} framework is problem- and survey-specific. The targets are set by the problem of interest, and the \textit{CL}s are designed to achieve this sole target. This suggests that conclusions drawn from applying the \textit{MCCL} approach should not be readily applied to different problems. For example, in this paper our goal is the measurement of shear one-point functions. Therefore, the results presented in this work are not appropriate to answer questions regarding two-point measurements of shear (e.g. spatial correlation of the shear measurements). A new \textit{MCCL} framework with different target values and diagnostics will need to be designed for each particular question.

\section{The Dark Energy Survey}\label{sec:data}
DES is a wide-field optical imaging survey that will cover 5000 deg$^{2}$ in the Southern Sky during its 5 years of operation and will record information of over 300 million galaxies. The survey area overlaps with other surveys such as the South Pole Telescope\footnote{http://pole.uchicago.edu/} (SPT) and the Visible and Infrared Survey Telescope for Astronomy (VISTA) Hemisphere Survey\footnote{http://www.vista-vhs.org/} (VHS). Focusing on Type Ia Supernovae, Baryon Acoustic Oscillations, galaxy clusters, and weak gravitational lensing as main cosmological probes, DES aims to study the nature of dark energy. The instrument achieved first light on 2012 September 12 and the main science survey officially started on 2013 August 12.

Images are taken with the Dark Energy Camera \citep[DECam; ][]{Flaugher:2012aa}, designed specifically for DES. The camera is mounted on the Blanco-4m telescope at Cerro Tololo Inter-American Observatory in Chile. DECam provides $0.27''$/pixel resolution. Good seeing on this site ranges between $0.7''$ and $1.1''$.

In this work, we will test our method using a fraction of the \textit{SV-A1} release, which covers $\sim$200 deg$^{2}$. Single exposures that were stacked to coadded images (straight averages), whose raw data are publicly available, were processed by the DES Data Management pipeline version ``SVA1'' (Yanny et al., in preparation). The images were taken during the SV period, which lasted between 2012 November and 2013 February. For this work, we selected images covering $\sim$50 deg$^{2}$ in the SPT-E field that are free of significant image artifacts. We demonstrate our \textit{MCCL} method on one image with an area of $\sim$0.5 deg$^{2}$, \textit{DES0441-4414}, while using the rest of this SPT-E subsample to derive the statistical errors. The area is sufficiently large and contains enough stars and galaxies for the simulations to be calibrated to this image.

\section{Ultra Fast Image Generator (UFig)}\label{sec:UFig}
In this paper we analyze images simulated with \textsc{UFig} \citep[][henceforth B13]{Berge:2013aa}. The image generation process consists of two steps. First, galaxy and star catalogs are generated. Then, the catalogs are turned into a coadded image. A brief overview of the properties of the \textsc{UFig}-generated galaxy catalogs, the PSF and noise models, and the shear field is given below, while a full description can be found in B13. Note that some of the models used in B13 are not fully realistic, but they provide a good starting point. The output from our \textit{MCCL} framework would inform us if more sophisticated models are needed to describe the data.

The \textit{MCCL} approach typically requires the simulation and analysis of many thousands of images. Thus, speed is crucial. In order not to be dominated by the image generation, its speed needs to be at least comparable to the analysis. Due to several computational optimizations, \textsc{UFig} is orders of magnitude faster than publicly available image simulators. In terms of runtime, generating an image is comparable to executing {\sc SExtractor} \citep[][henceforth SE]{Bertin:1996aa} on the same image, which sets the time scale in the \textit{MCCL} framework.

A key property of \textsc{UFig} is its flexibility in adjusting to different telescope setups. In this paper we choose to model \textit{r}-band coadded images taken by DECam, but it is straightforward to simulate images from other wide-field imaging surveys.

\subsection{Galaxies}\label{sec:ufiggalaxies}
A galaxy is simulated in \textsc{UFig} by sampling its light distribution photon-by-photon, and is then placed with a uniform probability on the image. Due to the finite number of photons sampled, the simulated galaxy images naturally include Poisson noise. PSF convolution in this approach is simply a displacement of the photons drawn from a probability distribution in the shape of the PSF (see Section~\ref{sec:ufigpsf}). We model the galaxy light distribution with a single-S\'{e}rsic profile to which we apply a distortion to generate the apparent ellipticity.

The radial profile is defined by a S\'{e}rsic index, an intrinsic magnitude and an intrinsic size. The latter two are non-trivially correlated. We parametrize and sample this distribution in a space, where magnitudes and sizes are approximately uncorrelated ($mag_{r}$, $\mathrm{log} \ r_{50,r}$). It is related to the magnitude and size plane ($mag$, $\mathrm{log} \ r_{50,i}$) through a rotation by an angle $\theta$ around a pivotal point ($mag_{p}$, $\mathrm{log} \ r_{50,p}$), i.e.
\begin{equation}\label{eq:rotation}
    \begin{pmatrix}
        mag \\
        \mathrm{log} \ r_{50,i}
    \end{pmatrix} =
    \begin{pmatrix}
        \mathrm{cos} \ \theta & \mathrm{sin} \ \theta\\
        -\mathrm{sin} \ \theta & \mathrm{cos} \ \theta
    \end{pmatrix}
    \begin{pmatrix}
        mag_{r} \\
        \mathrm{log} \ r_{50,r}
    \end{pmatrix}
    +
    \begin{pmatrix}
        mag_{p} \\
        \mathrm{log} \ r_{50,p}
    \end{pmatrix}.
\end{equation}
We parametrize the distribution of rotated galaxy intrinsic half-light radii $r_{50,r}$ with a log-normal distribution with rms dispersion $\sigma$. The distribution of rotated magnitudes $mag_{r}$ is approximated by the distribution of intrinsic magnitudes $mag$ shifted by $mag_{p}$, which was compiled by B13 from different ground- and space-based surveys. In other words, we assume the cumulated magnitude distribution to be approximately invariant under the rotation described in Eq.~\ref{eq:rotation}. This is a good approximation for small values of the rotation angle $\theta$. The two parameters $\theta$ and $\sigma$, the compiled cumulated magnitude distribution, and the pivotal point uniquely describe the two-dimensional distribution in the magnitude-size plane for our modeled galaxy sample. The S\'{e}rsic index distribution was derived by fitting single-S\'{e}rsic profiles to different galaxy samples in B13.

The intrinsic galaxy ellipticities are defined by \citep[see e.g.][]{Rhodes:2000aa}
\begin{equation}\label{eq:KSBmoments}
    e = e_{1} + ie_{2} = \frac{I_{11}-I_{22}+2iI_{12}}{I_{11}+I_{22}},
\end{equation}
where $I_{ij}$ are the unweighted quadrupole moments of the galaxy's light profile and $e_{1}$, $e_{2}$ are the two components of the ellipticity. In this paper, we sample $e_{1}$ and $e_{2}$ separately from normal distributions with mean zero and rms dispersion $e_{1,rms}$ and $e_{2,rms}$.

\subsection{Stars}\label{sec:ufigstars}
Since stars are typically brighter than galaxies, it is optimal to simulate them pixel-by-pixel rather than photon-by-photon. They are simulated directly on the image pixel grid and also placed on the image with a uniform probability. The profile is given by the PSF integrated within each pixel of the image grid (see Section~\ref{sec:ufigpsf}). Poisson noise is included by drawing a value from the corresponding Poisson distribution in every pixel.

To simulate a star, only a magnitude needs to be drawn. We sample a cumulated magnitude distribution derived from the stellar population synthesis model Besan\c{c}on \citep{Robin:2003aa}. In case the resulting intensity in a pixel is larger than DECam's saturation threshold, bleeding is modeled. 

\subsection{PSF}\label{sec:ufigpsf}
In this initial implementation of the \textit{MCCL} framework we choose as a baseline for future work a spatially constant, elliptical Moffat profile to describe the PSF. The elliptical Moffat profile can be derived from a two-dimensional linear transformation of the circular Moffat profile, which is given by \citep{Moffat:1969aa}
\begin{equation}\label{eq:moffat}
    I(r) = \frac{I_{0}}{\left(1+\left(\frac{r}{\alpha}\right)^{2}\right)^{\beta}},
\end{equation}
where the scale parameter $\alpha$ is related to the seeing. The profile is defined by the seeing, the exponent $\beta$, and the ellipticities $e_{1}$ and $e_{2}$. We find that the radial profile of stars in coadded DES images roughly follow a Moffat distribution with some variation in the parameters $\alpha$ and $\beta$.

In this initial implementation we choose for simplicity to fit a spatially invariant PSF to the image of interest (\textit{DES0441-4414}, see Section~\ref{sec:data}) in a pre-calibration step. We use in this work a PSF of FWHM $1.09''$, ellipticity $e_{1}=0.035$, $e_{2}=0.02$, and $\beta=3.5$ to match the mean PSF of this image. Note that this PSF size is slightly larger than the projected median seeing of the main survey ($\sim$$0.9''$). As shear measurement is more challenging with larger PSF sizes, we expect our \textit{MCCL} framework to produce results similar or better on an image with better seeing conditions.

\subsection{Noise}\label{sec:ufignoise}
Two different components add to the noise in \textsc{UFig}. First, we simulate galaxies down to magnitudes $r\sim29$. Since most of these faint galaxies are not detected, every image contains sky noise arising from many unresolved, faint galaxies. Second, we add a Gaussian background noise centered around 0 with a constant rms dispersion $\sigma_{N}$ across the image. This should capture noise induced by emission from the sky, and noise induced by the data processing. We perform furthermore Lanczos resampling \citep{Duchon:1979aa} with a kernel of width five pixels and a half-a-pixel offset on the simulated pixel grid. It allows us to mimic correlated noise in real images, while bypassing the expensive simulation and data reduction of raw images (see B13).

\subsection{Shear field}\label{sec:ufigshear}
We employ the following shear conventions in \textsc{UFig} and throughout this paper \citep{Rhodes:2001aa,Bartelmann:2001aa}
\begin{equation}
    \gamma_{1} = \frac{1}{2}\left(\partial_{1}^{2}-\partial_{2}^{2}\right)\Psi \ \ \ \mathrm{and} \ \ \ \gamma_{2} = \partial_{1}\partial_{2}\Psi,
\end{equation}
where $\Psi$ is the projected lensing potential.

To be close to real surveys, we use a $\Lambda$CDM shear power spectrum and model the shear field as a Gaussian random field. We choose $H_{0}=70 \ \mathrm{km \ s^{-1} \ Mpc^{-1}}$, $\Omega_{m}=0.3$, $\Omega_{\Lambda}=0.7$, $\sigma_{8}=0.8$. We simulate Gaussian random fields with \cite{Lang:2011aa}'s fast algorithm.

\section{Method}\label{sec:method}
The \textit{MCCL} framework is designed to validate the shear measurement process on simulated images and to test its robustness. RA13 identify three key iterative steps in the shear measurement process, which are labeled as \textit{Control Loops} (\textit{CL}), each with a distinct goal.

The first step (\textit{CL1}) is designed to find a fiducial configuration of simulation parameters such that the simulations agree with the data. In order to quantify the level of agreement, this step relies on defining a set of diagnostics and metric targets. The next step (\textit{CL2}) is to calibrate the shear measurement at this fiducial point. The final and computationally most demanding step (\textit{CL3}) aims to explore the parameter space volume for which data and simulations are in good agreement to ensure that the calibration scheme from \textit{CL2} is robust. This scheme is designed to ensure that the systematic errors on a given shear measurement are sub-dominant to the statistical errors. Should the results of \textit{CL3} show that the employed calibration scheme is not robust enough over all parameter space allowed by \textit{CL1}, then the whole \textit{MCCL} framework needs to be applied again with more stringent diagnostic requirements and possibly additional diagnostics.

It is clear now that in order to assess the robustness of this calibration scheme the generation and analysis of many tens or even hundreds of thousands of images is required. From a computational viewpoint this is only feasible if every step is very fast. This echoes our statement in Section~\ref{sec:UFig} on the importance of using \textsc{UFig} as our main image simulation tool.

The detailed implementation of each of the \textit{CLs} is presented below.

\subsection{Control Loop 1}\label{sec:loop1}
To make statements about the consistency of data and simulations output, we analyze three distributions described below as our main diagnostics. To assess how likely it is that two different distributions of data and simulations could be different realizations of the same underlying model, we use a $\chi^{2}$-method. We apply appropriate cuts to the three diagnostic distributions and bin them. This allows us to compute $\chi^{2}$ for each diagnostic and combine them by adding them up. For a number of different diagnostic distributions $\mathrm{\#Diag}$, the total $\chi^{2}_{red}$ for two binned datasets of different sizes is given by \citep[e.g.][]{Press:2002aa}
\begin{equation}\label{eq:reducedChi}
    \chi^{2}_{red} = \frac{1}{\mathrm{\sum_{k}N_{k}}}\sum_{i=1}^{\mathrm{\#Diag}}\sum_{j=1}^{N_{i}}\frac{\left(\sqrt{g_{i}/f_{i}}f_{ij}-\sqrt{f_{i}/g_{i}}g_{ij}\right)^{2}}{\sigma_{d,ij}^{2}+\sigma_{s,ij}^{2}}.
\end{equation}
Here, for the i-th diagnostic distribution of the real (simulated) image, $f_{ij}$ ($g_{ij}$) is the number of counts in the j-th bin. $N_{i}$ bins is the number of bins for this diagnostic with counts $f_{ij}$ above a certain threshold, and $f_{i}$ ($g_{i}$) is the sum of all counts in those bins. $\sigma_{d,ij}$ and $\sigma_{s,ij}$ are the errors of the data and the simulation for the i-th diagnostic distribution within the j-th bin. The errors need to be estimated in the data and the simulations. For the data, we estimate them by computing the variance in those bins for all the images in the sample. For the simulations, we generate many different realizations of the same input model and compute the variances in every bin. For this $\chi^{2}$-method the variables in each bin should follow a Gaussian distribution. We therefore only include bins with at least 50 objects (about 36000 objects are detected in the real image). We find this to be a good approximation.

\begin{figure}
    \centering
    \includegraphics[width=\linewidth]{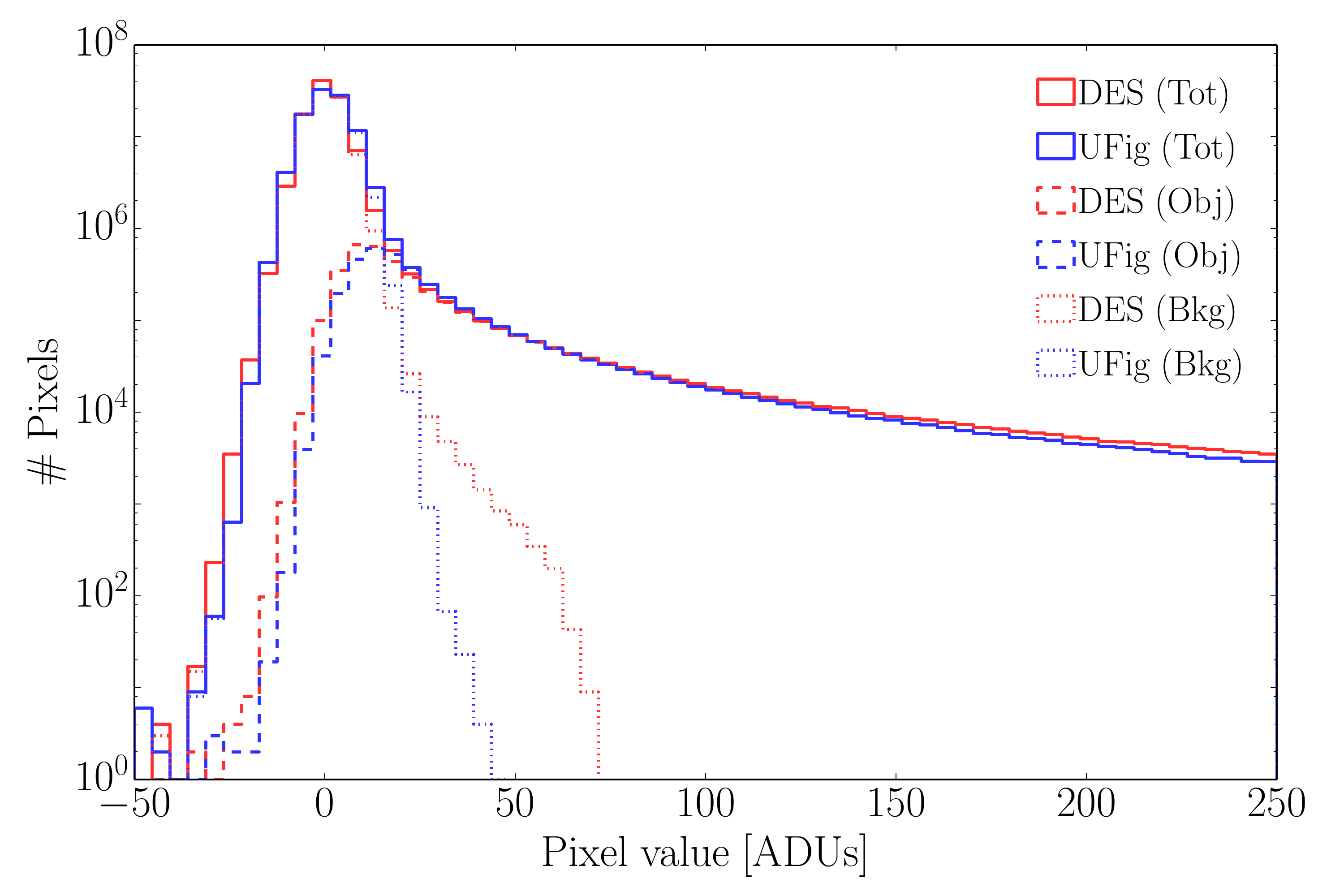}
    \caption{Histogram of the pixel values in ADUs for \textit{DES0441-4414} (red) and a simulated \textsc{UFig} image (blue) with the fiducial configuration after \textit{CL1}. The solid lines show the counts of all the pixels in the image. SE's segmentation map assigns pixels either to objects or the background. The dashed respectively dotted lines show the corresponding pixel counts.}
    \label{fig:diagPixels}
\end{figure}

\begin{figure*}
    \centering
    \includegraphics[width=\linewidth]{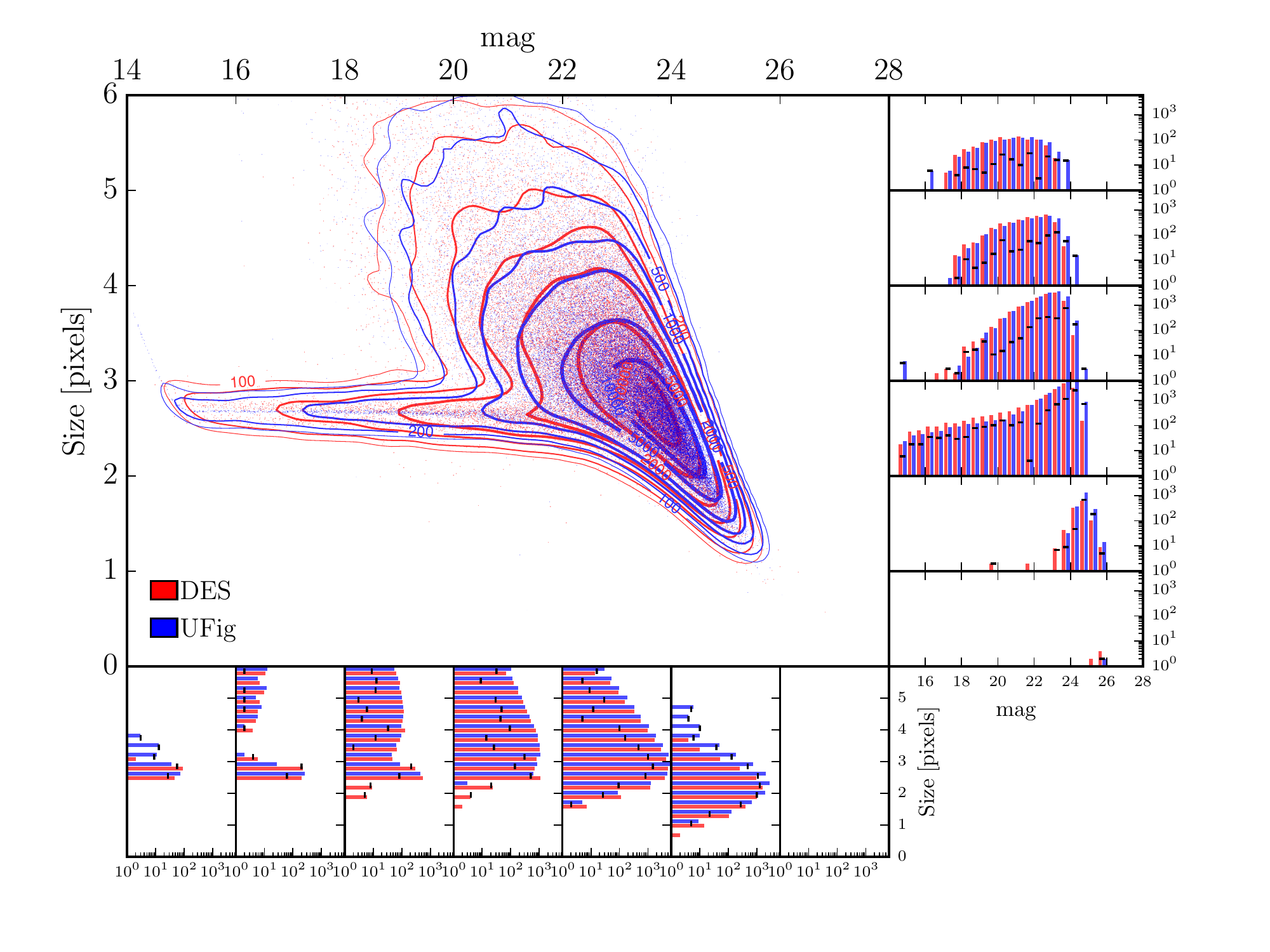}
    \caption[width=\linewidth]{Distribution of \textit{r}-band magnitudes (\textsc{MAG\_BEST}) and the sizes in pixels (\textsc{FLUX\_RADIUS}) of objects identified by SE. Isodensity contours of the number of objects track the shape of the distribution. Red is the \textit{DES0441-4414} and blue is a simulated \textsc{UFig} image with the fiducial configuration after \textit{CL1}. Histograms on the right and the bottom show the projected distributions in different size and magnitude bins. The black marks denote the difference between the red and blue histograms in every bin.}
    \label{fig:diagMagSize}
\end{figure*}

\begin{figure*}
    \centering
    \includegraphics[width=\linewidth]{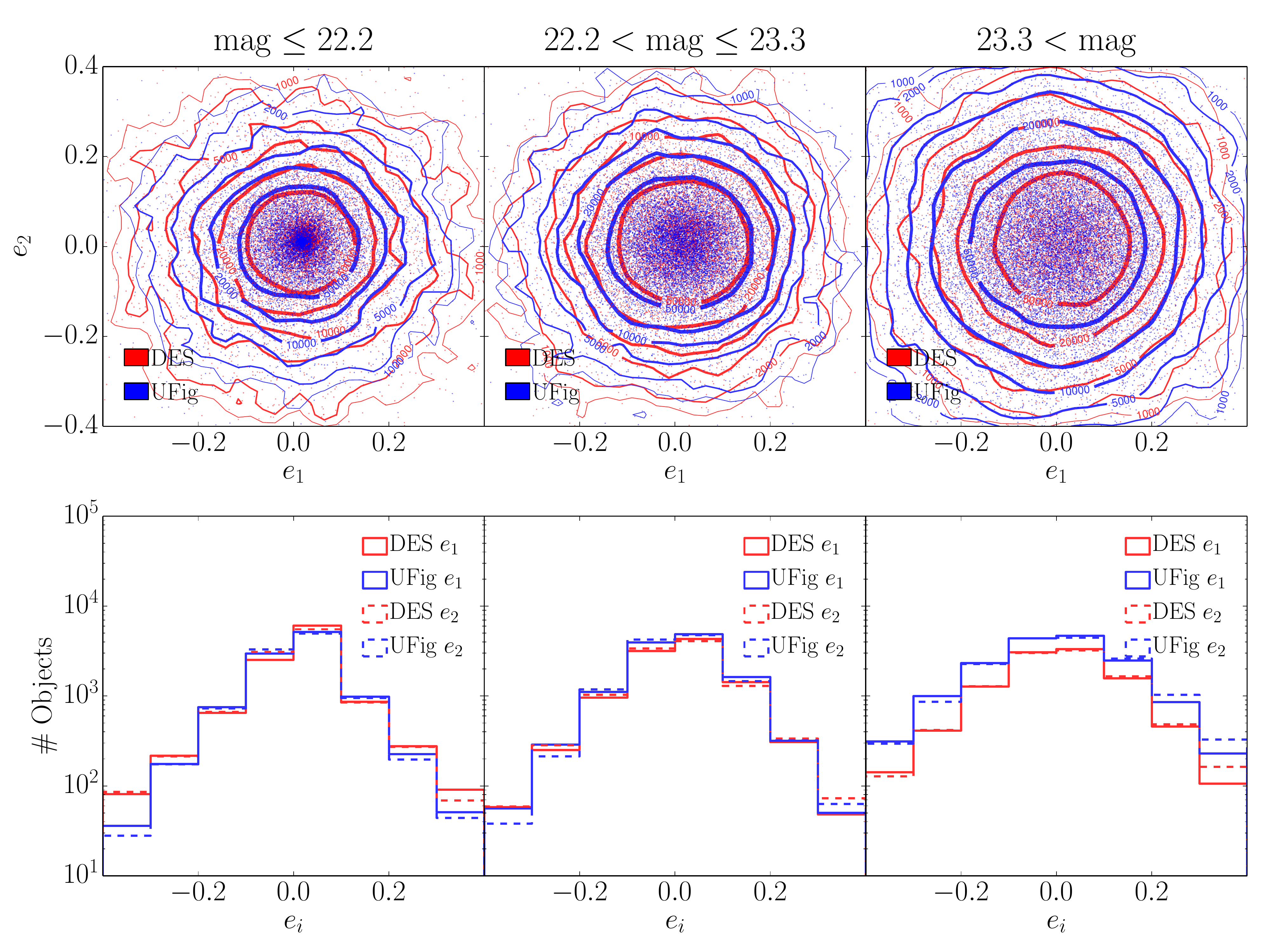}
    \caption{Ellipticity distributions for objects identified by SE in \textit{DES0441-4414} (red) and a simulated \textsc{UFig} image with the fiducial configuration after \textit{CL1} (blue) are shown. The objects are split up into three different \textit{r}-band magnitude (\textsc{MAG\_BEST}) bins, such that they contain approximately the same number of objects. In every bin, isodensity contours of identified objects in the ellipticity plane (see Eq. \ref{eq:KSBmoments}) (top row) and the corresponding histograms (bottom row) are shown.}
    \label{fig:diagEllip}
\end{figure*}

For this first implementation, we choose three diagnostics to break the degeneracies between the parameters we vary. They are refined iteratively to meet the requirements to pass \textit{CL3} (see Section \ref{sec:loop3}). Two of the three diagnostics use SE estimators. A comparison of the performance of certain estimators on these images is shown in \citep[e.g.][and references therein]{Bertin:1996aa,Chang:2014aa}. The three diagnostics are:
\begin{itemize}
\item Histogram of pixel values in ADUs (Fig. \ref{fig:diagPixels}): 1D

This is a valuable diagnostic to test the background properties of the image by comparing the peak of the distribution in the sky-subtracted images. Furthermore, it allows us to test the magnitude zero point of the image. Different magnitude zero points shift the tail of the large pixel values vertically, as they affect the number of pixels with small respectively large pixel values.

\item Binned magnitude versus size-plane (Fig. \ref{fig:diagMagSize}): 2D

This diagnostic probes the magnitude and size distributions of identified objects in the images and their correlation. We use the SE columns \textsc{MAG\_BEST} for the magnitude and \textsc{FLUX\_RADIUS} for the size.

\item Binned $e_{1}$ versus $e_{2}$-plane in three different magnitude bins (Fig. \ref{fig:diagEllip}): 2D

This tests the ellipticity distribution of identified objects. We estimate the ellipticity using a version of Eq. \ref{eq:KSBmoments} with weighted quadrupole moments. We split the objects up into three different magnitude bins, each containing a similar number of objects. This allows us to probe the ellipticity distribution in each S/N bin individually. With the high-S/N bin being least affected by the effects of the PSF, different intrinsic ellipticity distributions can be distinguished. The low-S/N bin, which contains faint objects whose shape is dominated by the PSF, on the other hand allows us to test the properties of the PSF.
\end{itemize}

We minimize $\chi^{2}_{red}$ to find a fiducial configuration. For this first implementation, we choose to vary six simulation parameters to generate new samples describing the galaxy population, the image properties, and the noise level: the magnitude zero point of the image $mag_{0}$, the rms of the log-normal size distribution $\sigma$ (Section~\ref{sec:ufiggalaxies}), the rotation angle $\theta$ between the magnitude and size plane and the plane where the quantities are approximately uncorrelated (Section~\ref{sec:ufiggalaxies}), the rms of the Gaussian background noise $\sigma_{N}$ (Section~\ref{sec:ufignoise}), and the rms of the Gaussian distributions for the ellipticities $e_{1,rms}$ and $e_{2,rms}$ (Eq.~\ref{eq:KSBmoments}). Those six parameters are not constrained by fits performed in B13. For each configuration an image is simulated and the $\chi^{2}_{red}$-value is computed (Eq.~\ref{eq:reducedChi}).

The $\chi^{2}_{red}$ minimization procedure is designed to find a sensible parameter regime in a small number of iterations. It consists of two steps: First, we sample the parameter space coarsely and identify the region in the parameter space where the minimum $\chi^{2}_{red}$ is located. Then, by successive one-dimensional minimizations we find the minimum $\chi^{2}_{red}$. We vary each parameter while holding the others fixed and compute the new $\chi^{2}_{red}$ values. The specific parameter value that minimizes $\chi^{2}_{red}$ defines a new configuration. We repeat this step iteratively until it converges. The final result of the iteration is the fiducial configuration that is analyzed in the subsequent \textit{CL}s.

If Eq.~\ref{eq:reducedChi} can be applied, i.e. the quantities in each bin of the diagnostic distributions are Gaussian distributed, then confidence limits on the parameters can be computed. In this case, for a model with six degrees of freedom the 95\% confidence limits are given by \citep[e.g.][]{Chernick:2003aa}
\begin{equation}\label{eq:chiSquareRange}
    \Delta\chi^{2}_{red} = \frac{1}{\mathrm{\sum_{k}N_{k}}}\cdot 12.59.
\end{equation}
This gives for every parameter a range of values for which data and simulations are statistically consistent (see Appendix).

\begin{figure*}
    \centering
    \includegraphics[width=\linewidth]{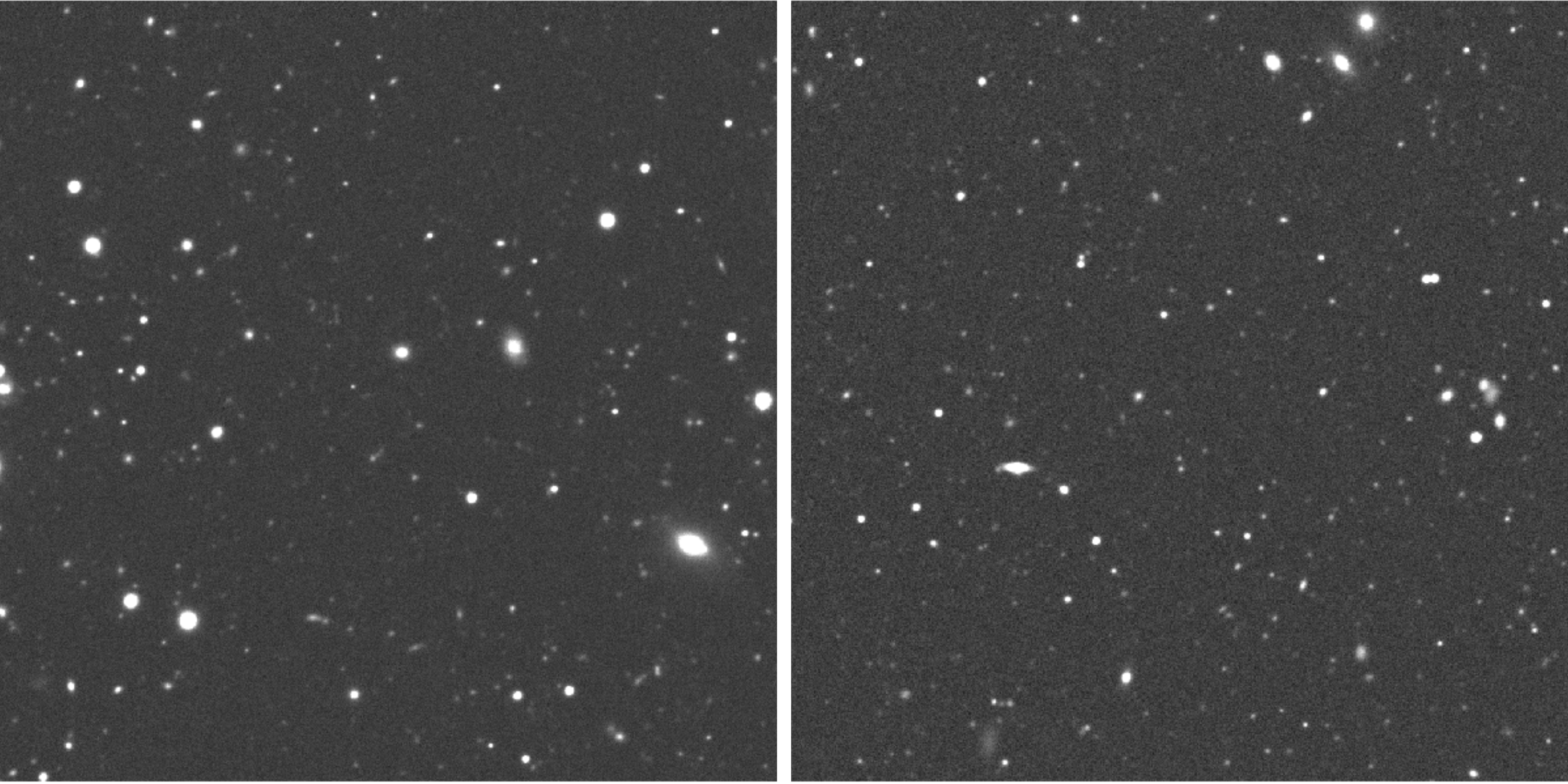}
    \caption[width=\linewidth]{Comparison of a DES SV image (\textit{DES0441-4414}; left) and a \textsc{UFig} simulated image after \textit{CL1} (right). A 4 arcmin$^{2}$ segment of the total 0.5 deg$^{2}$ images is shown. The same color scale has been applied to both images.}
    \label{fig:images}
\end{figure*}

\subsection{Control Loop 2}\label{sec:loop2}
The task of \textit{CL2} is to calibrate the shear measurement by comparing input and estimated shear signal on simulated data. The image that we use to illustrate the \textit{MCCL} framework is part of the DES \textit{SV-A1} release, which covers about 200 deg$^{2}$. The larger the area simulated for calibration is, which needs to be larger than the size of the dataset, the more precise is the resulting calibration scheme. On the other hand however the computational costs increase for a larger area. We choose to simulate an area equivalent to 1000 deg$^{2}$ for any given configuration. On the galaxies detected in those images we apply a S/N-cut of 15, where we define the S/N as SE's \textsc{FLUX\_BEST}/\textsc{FLUXERR\_BEST}, and a size-cut of 1.2 times the PSF size using SE's \textsc{FLUX\_RADIUS} measurement. This allows us to select galaxies large and bright enough for calibration.

We follow \cite{Rhodes:2001aa} to first order to estimate the galaxy shear
\begin{equation}\label{eq:gammaEst}
    \hat{\boldsymbol{\gamma}} = \frac{e'}{2-\langle \left|e\right|^{2}\rangle},
\end{equation}
where
\begin{equation}\label{eq:ellipDef}
    e' = \frac{J'_{11}-J'_{22}+2iJ'_{12}}{J'_{11}+J'_{22}}
\end{equation}
is the lensed ellipticity, and $e$ the unlensed one. In the weak lensing limit we can approximate
\begin{equation}\label{eq:ellipApprox}
    \langle \left|e\right|^{2}\rangle \approx \langle \left|e'\right|^{2}\rangle.
\end{equation}
We use SE's \textsc{X2WIN\_IMAGE}, \textsc{Y2WIN\_IMAGE}, and \textsc{XYWIN\_IMAGE} to measure the weighted quadrupole moments of the PSF-convolved image $\tilde{J}'_{ij}$. To linear order and ignoring weight function terms, the PSF can approximately be corrected for using
\begin{equation}\label{eq:PSFcorr}
    J'_{ij} = \tilde{J}'_{ij} - P_{ij},
\end{equation}
where $P_{ij}$ is the mean of the weighted quadrupole moments of the stars. We use $J'_{ij} = \tilde{J}'_{ij}$ in Eq. \ref{eq:ellipDef} when PSF correction is not applied.

The galaxies are then binned in input shear signal and the mean estimated shear is computed in every bin. We calibrate the shear measurement to first order by fitting and applying a linear correction
\begin{equation}\label{eq:correction}
    \hat{\gamma}_{i} = \alpha_{i}\gamma_{in,i} + \beta_{i},
\end{equation}
where $\gamma_{in,i}$ is the input shear.

\subsection{Control Loop 3}\label{sec:loop3}
Knowing the ranges of parameter values for which data and simulations are statistically consistent from \textit{CL1}, we can test the robustness of the calibration schemes for shear measurements for different configurations in this parameter space volume (\textit{CL3.1}). We vary each parameter in a range slightly larger than that allowed by the data, while keeping the other parameters fixed, and calibrate the shear measurement on this new location in parameter space (see Section~\ref{sec:loop2}). We then explore by which amount the calibration ($\alpha$ and $\beta$; see Eq. \ref{eq:correction}) changes relative to the calibration on the fiducial configuration resulting from applying \textit{CL1}. This uncertainty in the shear calibration corresponds to the systematic error we expect in the shear measurement.

The resulting multiplicative and additive biases (Eq. \ref{eq:defMultAddBias}) due to the uncertainty in the fiducial configuration are computed by evaluating
\begin{equation}\label{eq:multAddBiasAlphaBeta}
    m_{i} = \frac{\Delta\alpha_{i}}{\alpha_{i}} \ \ \ \mathrm{and} \ \ \ c_{i} = \frac{\Delta\beta_{i}}{\alpha_{i}},
\end{equation}
where $\Delta\alpha_{i}$ and $\Delta\beta_{i}$ are the changes relative to the fiducial calibration parameters. We require $m_{i}$ and $c_{i}$ to meet the targets set in Section~\ref{sec:shearMsrtProb}, otherwise the diagnostics themselves need to be refined and additional tests could be required (\textit{CL3.2}), affecting all the previous loops.

\section{Results}\label{sec:results}

\subsection{Control Loop 1}\label{sec:cl1}
Excerpts of the \textit{DES0441-4414} image and the \textsc{UFig} image simulated with the fiducial configuration are displayed in Fig. \ref{fig:images}. They apper similar visually. For a quantitative comparison, Figures \ref{fig:diagPixels}-\ref{fig:diagEllip} show the diagnostic plots for the DES image and the \textsc{UFig} image. The combined $\chi^{2}_{red}$ of the individual values for each diagnostic has a value of 1.06. Thus, the fiducial configuration we find is a good fit to the data in the chosen diagnostics. To avoid combining very different $\chi^{2}_{red}$ values, we assure that the individual $\chi^{2}_{red}$ values are also close to 1. For the fiducial configuration, the individual ones for each diagnostic are within $\left|\chi^{2}_{red}-1\right|<0.4$ (see Appendix). By varying the binning scheme we have checked that we recover similar fiducial configurations and confidence limits.

Fig. \ref{fig:diagPixels} shows the histograms of pixel values for all the pixels in both images (solid). The overall behavior agrees well ($\chi^{2}_{red} \approx 1.38$). The histograms agree well around the peak, with the distribution of the pixels in the \textsc{UFig} image being slightly broader. The pixels are furthermore divided using SE's segmentation map into two sets to allow us to understand differences and similarities better. One set contains all the pixels associated with identified objects (dashed), and the other those associated with the background (dotted). The histograms of pixels associated with objects agree well ($\chi^{2}_{red} \approx 1.10$). We however observe a low-level discrepancy in the background pixel histograms at high pixel values. While our noise model including Gaussian noise in every pixels seems to be a good approximation around the peak of the histogram, it does not account for the background pixels with larger positive pixel values. As the number of background pixels is small compared to the total number of pixels with pixel values of $\gtrsim$ 30 ADUs, those differences do not affect the value of $\chi^{2}_{red}$ significantly.

Fig. \ref{fig:diagMagSize} displays the magnitude-size plane of objects identified by SE in both the simulation and the data. Overall, the distributions resemble each other qualitatively and quantitatively ($\chi^{2}_{red} \approx 1.26$). In particular, the main bulk of the galaxy distributions, the location of the stellar loci, and the saturation turnoffs all agree well. Some slight differences can however be noted. The dispersion around the stellar locus is larger in the DES image, which is due to our simple PSF model constant in size. Furthermore, the shapes of the density contour lines and the magnitude limits are slightly different. We believe that changes in the galaxy model would improve this.

The different magnitude limits and the discrepancies in the background-only histograms of pixel values call for more noise in the simulations. Increasing the width of the Gaussian background would on the other hand however aggravate the discrepancy around the background peak. To resolve this tension (see Appendix), a more sophisticated background model easing some of the simplifying assumptions on the properties of the background is needed \citep[for an overview of possible extensions see][]{Rowe:2014aa}. An analysis of the two-point correlation function will reveal structures in the background not yet modeled and will serve as an additional diagnostic.

The ellipticity planes in the different magnitude bins are shown in Fig. \ref{fig:diagEllip}. Due to the ellipticity introduced by the PSF, the mean of the $e_{i}$-distributions is shifted towards positive values and thus there is a small asymmetry. Note that the galaxies we include in the calibration of the shear measurement are mainly in the two brighter magnitude bins where the distributions match well ($\chi^{2}_{red} \approx 1.35$ and $\chi^{2}_{red} \approx 0.63$). In the brightest magnitude bin, the distributions deviate slightly for values of $|e_{i}|>0.3$. We believe this is caused by our choice of the intrinsic ellipticity distributions being normal in $e_{i}$ (see Eq. \ref{eq:KSBmoments}). Changes in the intrinsic ellipticity distribution can improve the agreement between the data and the simulation. In the faintest magnitude bin, the distributions do not match well ($\chi^{2}_{red} \approx 0.21$). As noted above, there seem to be more faint objects detected in the \textsc{UFig} image (about $\sim15\%$ more detected objects in total). Furthermore, there are differences between the ellipticity distributions in this bin. This can be attributed to the simple PSF model we choose, as the objects in this bin are mostly dominated by the PSF. As we apply a S/N-cut of 15 (corresponds to $mag\sim23$), differences in the faintest magnitude are potentially not relevant for the calibration of the shear measurement. Nevertheless, it is only by looking at the results of a future, more rigorous \textit{MCCL} analysis including parameters describing the PSF model that we can assess whether the differences in the faintest magnitude bin are relevant for shear measurement.

\begin{deluxetable}{ccc}
  \tabletypesize{\textwidth}
  \tablecolumns{3}
  \tablewidth{0pt}
  \tablecaption{Calibration coefficients $\alpha_{i}$ and $\beta_{i}$ for the fiducial configuration} 
  \tablehead{\colhead{Coefficient} & \colhead{PSF-uncorrected} & \colhead{PSF-corrected}}
  \startdata
    $\alpha_{1}$ & $0.199 \pm 0.001$ & $0.380 \pm 0.002$ \\
    $\alpha_{2}$ & $0.197 \pm 0.001$ & $0.377 \pm 0.002$ \\
    $\beta_{1}$ & $(4.40 \pm 0.01)\cdot10^{-3}$ & $(0.82 \pm 0.02)\cdot10^{-3}$ \\
    $\beta_{2}$ & $(2.86 \pm 0.01)\cdot10^{-3}$ & $(1.10 \pm 0.03)\cdot10^{-3}$ \\
  \enddata
  \label{tab:coeffValues}
\end{deluxetable}

\begin{figure*}
    \centering
    \includegraphics[width=.85\linewidth]{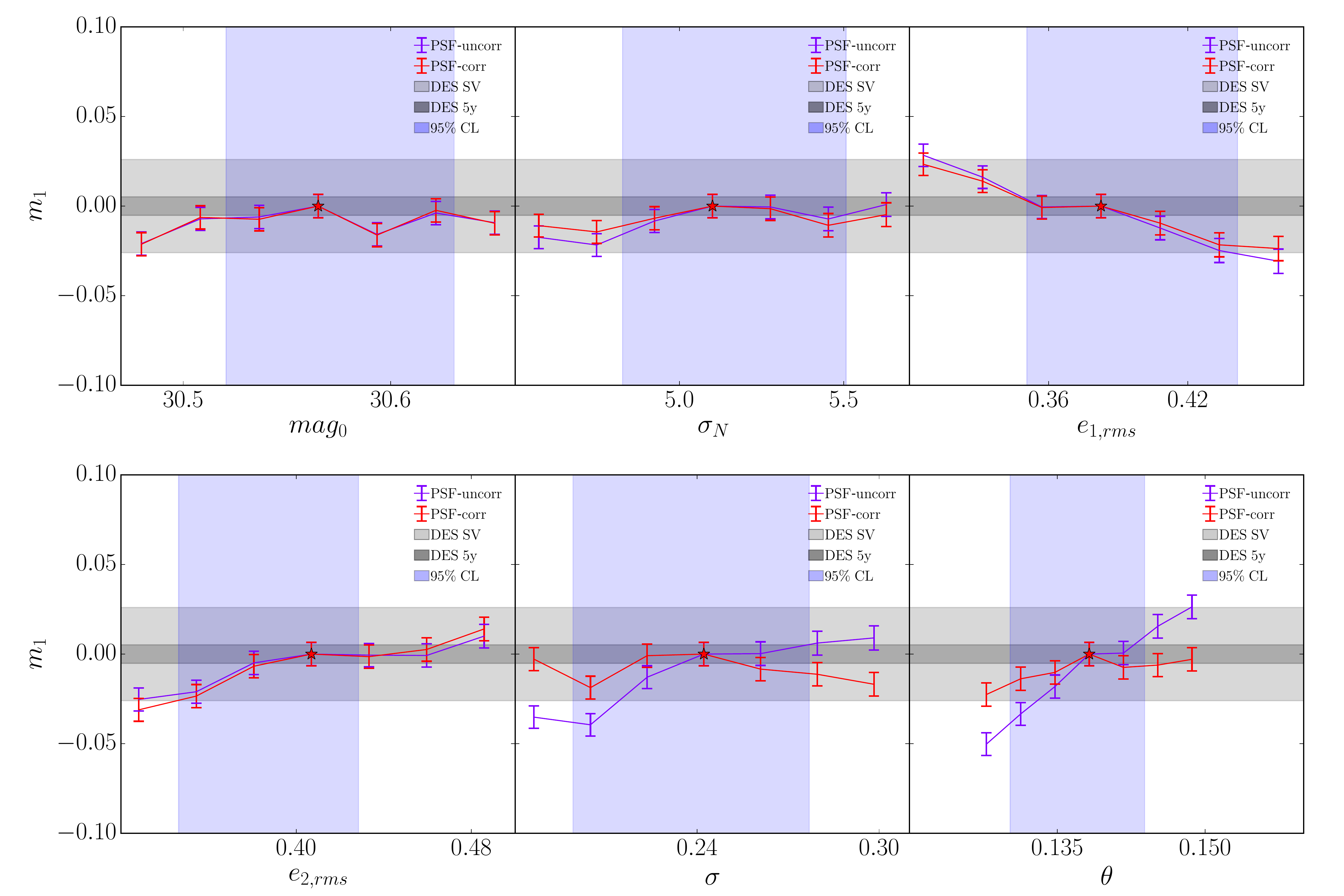}
    \caption[width=\linewidth]{Multiplicative bias in the measurement of $\gamma_{1}$ as a function of different parameter values and different shear measurement methods. We simulate images equivalent to an area of 1000 deg$^{2}$ for every configuration to calibrate the shear measurement. The change is relative to the central data point, our fiducial shear calibration. The vertical blue bands show the range in parameter values data and simulations are statistically consistent (95\% confidence limits) (see Section~\ref{sec:ufiggalaxies}). The horizontal gray bands correspond to the required accuracy in the shear measurement of a 200 deg$^{2}$ (light gray) and 5000 deg$^{2}$ survey for the measurement not to be systematics-limited. The star denotes the fiducial configuration.}
    \label{fig:resDeltaM1}
\end{figure*}
\begin{figure*}
    \centering
    \includegraphics[width=.85\linewidth]{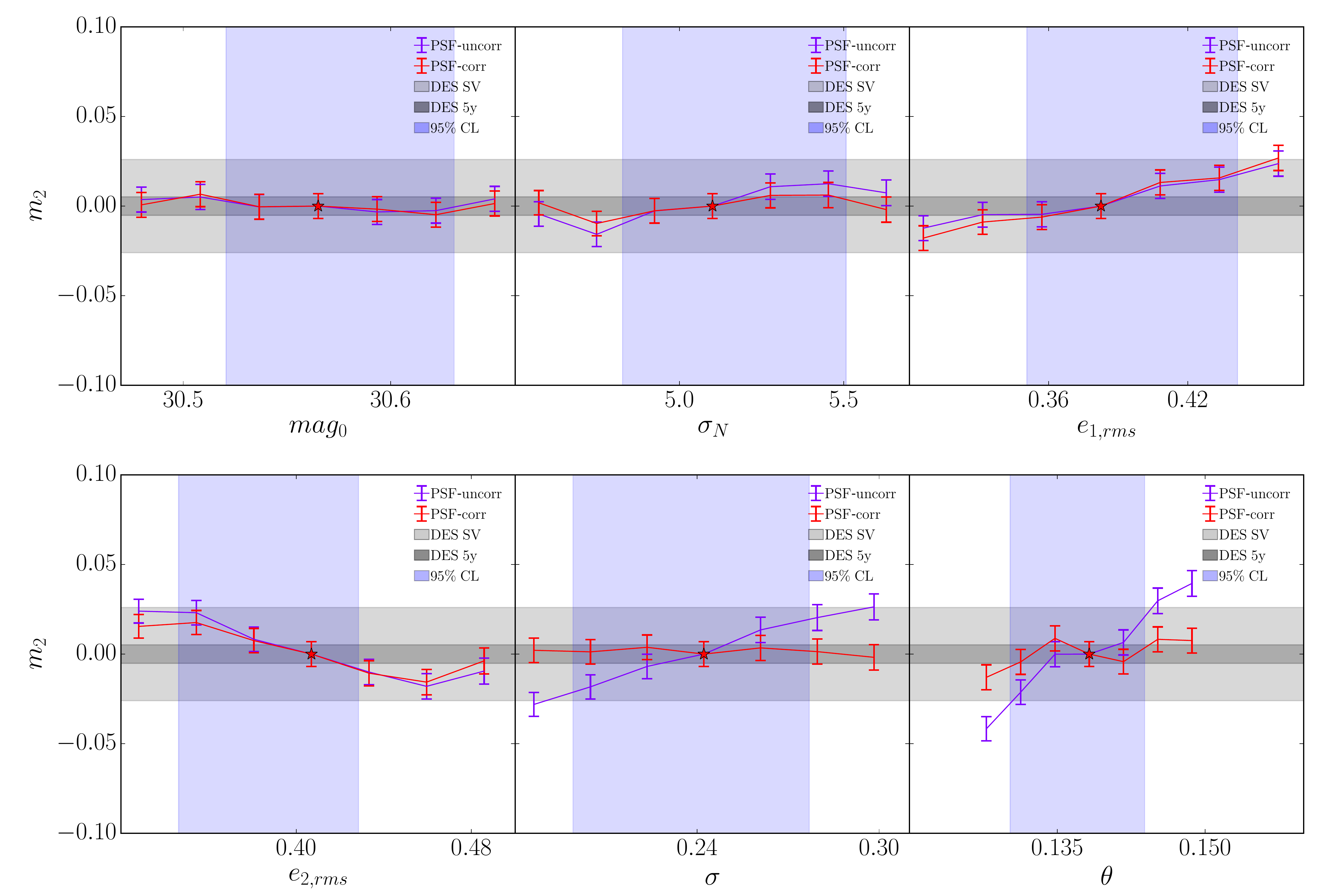}
    \caption[width=\linewidth]{Multiplicative bias in the measurement of $\gamma_{2}$. Similar to Fig. \ref{fig:resDeltaM1}.}
    \label{fig:resDeltaM2}
\end{figure*}
\begin{figure*}
    \centering
    \includegraphics[width=.85\linewidth]{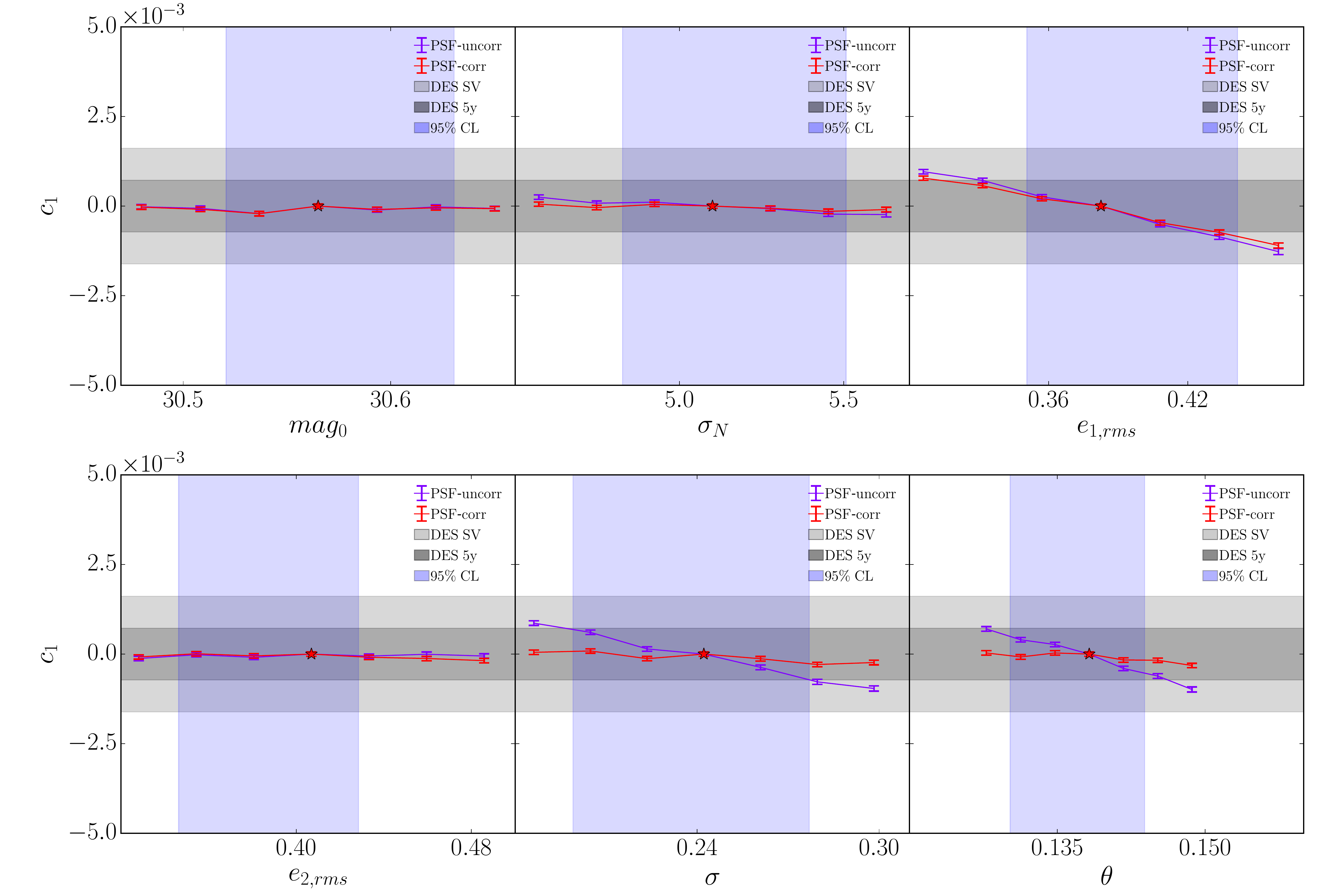}
    \caption[width=\linewidth]{Additive bias in the measurement of $\gamma_{1}$. Similar to Fig. \ref{fig:resDeltaM1}.}
    \label{fig:resDeltaC1}
\end{figure*}
\begin{figure*}
    \centering
    \includegraphics[width=.85\linewidth]{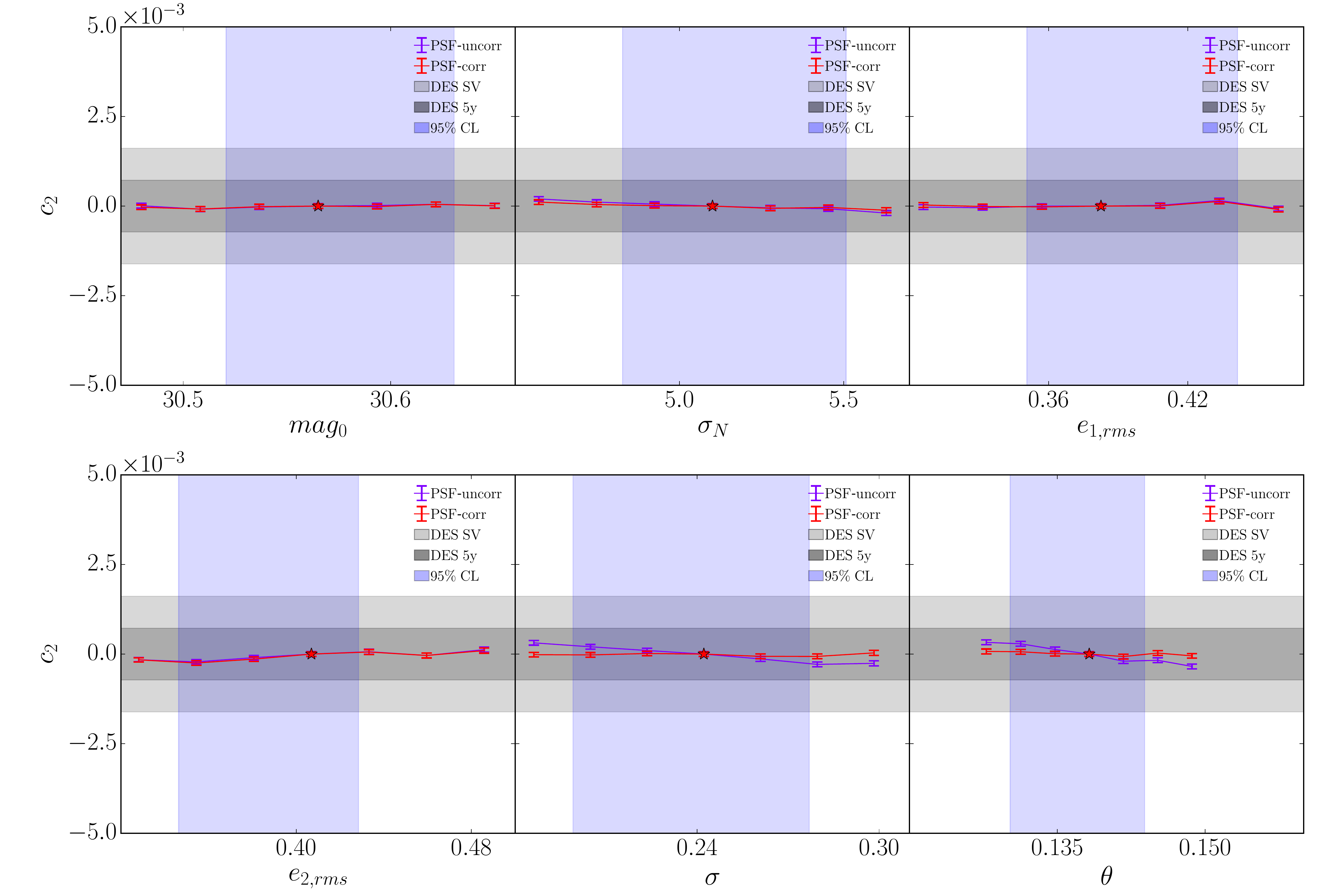}
    \caption[width=\linewidth]{Additive bias in the measurement of $\gamma_{2}$. Similar to Fig. \ref{fig:resDeltaC1}.}
    \label{fig:resDeltaC2}
\end{figure*}

\subsection{Control Loops 2 and 3}\label{sec:cl3}
We perform a tolerance analysis of the shear calibration, as described in Section~\ref{sec:loop3}. We vary the same six parameters as in Section~\ref{sec:loop1}, $mag_{0}$, $\sigma$, $\theta$, $\sigma_{N}$, $e_{1,rms}$, and $e_{2,rms}$. The allowed parameter ranges by the data are given by the analysis performed in Section~\ref{sec:cl1} (see Appendix). They correspond to the 95\% confidence limits we take as a measure of statistical consistency of different configurations. For each of these parameters, we compute the change in calibration relative to the fiducial model (see Table~\ref{tab:coeffValues}) at six different points around the fiducial configuration. For every data point we simulate an area of 1000 deg$^{2}$. Thus, to calibrate the shear measurement with the precision required for 200 deg$^{2}$ survey, we need to simulate 37000 deg$^{2}$.

Figures \ref{fig:resDeltaM1} and \ref{fig:resDeltaM2} show how uncertainties in the input parameters result in multiplicative biases. We use the two shape measures described by Equations \ref{eq:gammaEst}-\ref{eq:PSFcorr}. We find that, for the six parameters we vary, the PSF-corrected shape measurement calibrated through the \textit{MCCL} framework is robust enough for a DES SV-like 200 deg$^{2}$ survey in terms of the requirement described in Section~\ref{sec:shearMsrtProb}. As discussed in \citep{Refregier:2014aa} unknown systematics or effects not yet included in the simulations may affect the shear measurement. However, the \textit{MCCL} approach provides a framework for testing aspects of the measurement process that are in doubt. The PSF-uncorrected shape measurement does not perform as well as the PSF-corrected one, and lies slightly outside the tolerance band in some parameters. To make statements about whether the calibration scheme is robust enough for a 5-year DES-like 5000 deg$^{2}$-survey in the parameters varied, a larger area needs to be simulated to increase the accuracy of the calibration. Furthermore, as described in Section~\ref{sec:shearMsrtProb}, achieving this new target requires refinements on the \textit{MCCL} framework.

Figures \ref{fig:resDeltaC1} and \ref{fig:resDeltaC2} show the resulting additive biases. For the parameters considered, both shape measures already even satisfy the requirements for a full DES-like survey with 5 years worth of images.

We find in this first tolerance analysis that the calibration of the shear measurement seems to depend sensitively on the intrinsic ellipticity distribution. While there is not a significant additive bias due to an uncertainty in $e_{1,rms}$ and $e_{2,rms}$, the ellipticity distribution needs to be taken special care of such that no significant multiplicative bias is induced. The diagnostics likely need to be refined further to reduce this residual systematic effect such that stricter targets can be met in further \textit{MCCL} analyses.

\section{Conclusion}\label{sec:conclusion}
We have presented an initial implementation of the \textit{Monte Carlo Control Loops} \citep{Refregier:2014aa}, a novel approach for weak lensing shear measurements. The method contains a set of three \textit{Control Loops (CL)} applied to data and image simulations to forward-model the shear measurement process. They are designed specifically to calibrate the shear measurement and test its robustness with the goal of reaching a certain sensitivity. The requirements in this paper are chosen such that the lensing measurement, assuming spatially invariant systematic errors, on the final dataset of a DES- and also DES SV-like imaging survey is not limited by systematic errors, i.e. the systematic error of the measurement is smaller than the statistical error. 

The \textit{MCCL} approach provides a consistent way of analyzing systematic errors in the measurement. It allows us to probe potential sources of error for their effect on the measurement, e.g. noise bias \citep{Refregier:2012aa,Kacprzak:2012aa} and model bias \citep[e.g.][]{Kacprzak:2014aa}, provided that they are included in the simulations. However, the simulation and analysis of a large number of images is essential in this approach. To not be limited computationally, every step in the \textit{CL}s needs to be fast, especially the generation of images. This led to the development of the Ultra Fast Image Generator \citep{Berge:2013aa}, whose speed is comparable to executing {\sc SExtractor} \citep{Bertin:1996aa}, the image analysis tool used in this paper.

We present a first implementation of the \textit{MCCL} framework using an image taken during the Science Verification (SV) phase of DES. For this purpose, we choose a spatially invariant PSF model, vary six simulation parameters, and consider only one-point shear measurements. With these assumptions, we find that the image calibration achieves multiplicative and additive biases within the needed weak lensing precision for a DES SV-like (200 deg$^{2}$) survey, assuming them to be spatially invariant. We also find with the tolerance analysis that the shear measurement is very sensitive to the intrinsic ellipticity distribution. Furthermore, we find an interplay between the magnitude-size and the histogram of pixel values diagnostics in fitting the noise level to the image. To accommodate both diagnostics, an extension of the Gaussian noise model will be implemented in future work.

To achieve our goal of not being systematics-limited when measuring shear on a 5000 deg$^{2}$ DES-like survey, several features in the \textit{MCCL} framework need to be refined. First, we will incorporate more realistic instrument and noise model in the image simulations. Next, we like to extend this framework to include two-point functions in the analysis. In addition, we are planning to test the effects of a spatially varying PSF and other PSF models. We will also explore the effect of more complex galaxy models and non-uniform distributions of galaxies on the calibration of the shear measurement. And finally, a more rigorous tolerance analysis varying more simulation parameters is required.

The results we present in this work require the simulation of about 40000 deg$^{2}$ of images. From a simple extrapolation of this figure, the computational resources needed for the full 5-year DES data appear large. However, several improvements on the framework can readily result in significant speed-ups. For example, better sampling strategies can in principle speed up the tolerance analysis in \textit{CL3.1} by orders of magnitude, which is by far the most computationally expensive step in this work. Furthermore, improvements in the diagnostics used in \textit{CL1} will increase the discriminatory power between different simulation configurations, reducing further the parameter space one needs to sample, and thus the computational time.

All these improvements will pave the way to exploit the full potential of weak lensing through the understanding of systematic effects within the \textit{MCCL} framework.

\section{Acknowledgements}
The authors would like to thank Sarah Bridle, Tomasz Kacprzak, David Bacon, Matthew Becker, Barnaby Rowe, Gary Bernstein, and the other members of the DES collaborations for useful discussion. This work was supported in part by grants $200021\_14944$ and $200021\_143906$ from the Swiss National Science Foundation.

Funding for the DES Projects has been provided by the U.S. Department of Energy, the U.S. National Science Foundation, the Ministry of Science and Education of Spain, the Science and Technology Facilities Council of the United Kingdom, the Higher Education Funding Council for England, the National Center for Supercomputing Applications at the University of Illinois at Urbana-Champaign, the Kavli Institute of Cosmological Physics at the University of Chicago, Financiadora de Estudos e Projetos, Funda\c{c}\~{a}o Carlos Chagas Filho de Amparo \`{a} Pesquisa do Estado do Rio de Janeiro, Conselho Nacional de Desenvolvimento Cient\'{i}fico e Tecnol\'{o}gico and the Minist\'{e}rio da Ci\^{e}ncia e Tecnologia, the Deutsche Forschungsgemeinschaft and the Collaborating Institutions in the Dark Energy Survey. 

The Collaborating Institutions are Argonne National Laboratory, the University of California at Santa Cruz, the University of Cambridge, Centro de Investigaciones Energeticas, Medioambientales y Tecnologicas-Madrid, the University of Chicago, University College London, the DES-Brazil Consortium, the Eidgen\"{o}ssische Technische Hochschule (ETH) Z\"{u}rich, Fermi National Accelerator Laboratory, the University of Edinburgh, the University of Illinois at Urbana-Champaign, the Institut de Ciencies de l'Espai (IEEC/CSIC), the Institut de Fisica d'Altes Energies, Lawrence Berkeley National Laboratory, the Ludwig-Maximilians Universit\"{a}t and the associated Excellence Cluster Universe, the University of Michigan, the National Optical Astronomy Observatory, the University of Nottingham, The Ohio State University, the University of Pennsylvania, the University of Portsmouth, SLAC National Accelerator Laboratory, Stanford University, the University of Sussex, and Texas A\&M University.

\bibliographystyle{apj}
\bibliography{calibratedUfig}

\clearpage
\appendix

\section{Reduced $\chi^{2}$ as a function of different simulation parameters}

As described in Section~\ref{sec:loop1}, we search for the fiducial simulation configuration by minimizing the $\chi^{2}_{red}$ defined in Eq. \ref{eq:reducedChi}. In this Appendix, we describe in detail the minimization procedure and point out some features in the resulting $\chi_{red}^{2}$ functions, which may indicate interesting physical insights to the data.  

\begin{figure*}
    \centering
    \includegraphics[width=.9\linewidth]{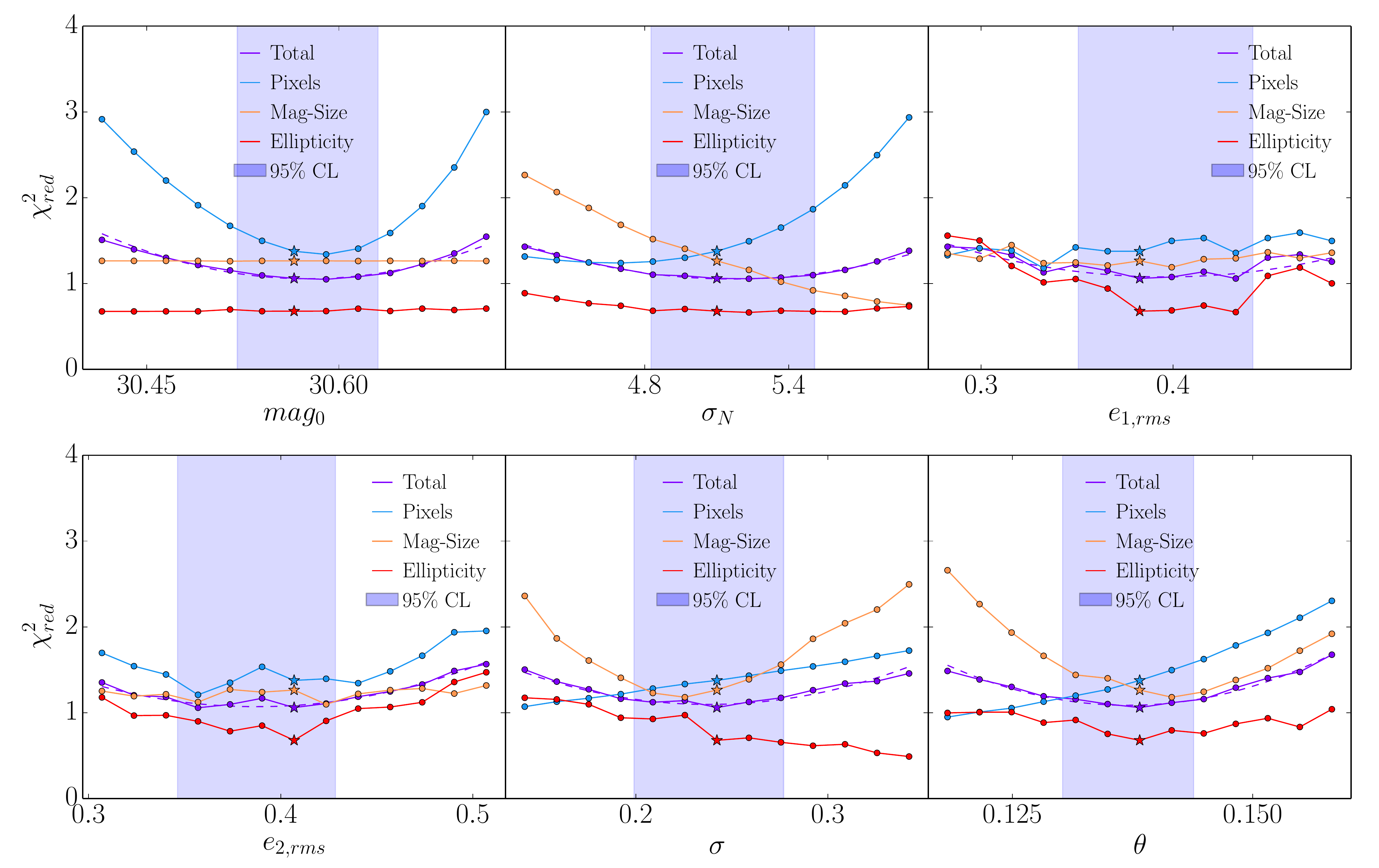}
    \caption{$\chi^{2}_{red}$-curves for the three different diagnostics as a function of different parameter values are shown. The star denotes the fiducial configuration. The vertical blue bands are computed with Eq. \ref{eq:chiSquareRange} and show the 95\%-confidence limits. They correspond to the blue bands in Figures \ref{fig:resDeltaM1}-\ref{fig:resDeltaC2}.}
    \label{fig:onlyChiLines}
\end{figure*}

For each of the six simulation parameters considered, we systematically vary its value around some initial guess and calculate $\chi^{2}_{red}$ while holding the other parameters fixed. The six parameter values that yield the minimum $\chi^{2}_{red}$ are then used for the next iteration and the process continues until it converges about the minimum. In this final set of parameters, the $\chi_{red}^{2}$ values along each of the one-dimensional axes are shown in Fig. \ref{fig:onlyChiLines}. The blue shaded bands, which correspond to the blue bands in Figures \ref{fig:resDeltaM1}-\ref{fig:resDeltaC2}, are the 95\%-confidence limits for each of the parameters (see Eq. \ref{eq:chiSquareRange}). Table \ref{tab:paramValues} lists the corresponding parameter values in the plots.

To better understand how the various diagnostics affect the resulting $\chi^{2}_{red}$-function, we split it up into the contributions of each diagnostic. For the fiducial configuration, which is denoted by a star, we find $|\chi^{2}_{red}-1|<0.4$ for all the individual diagnostics and their combined sum. The total $\chi^{2}_{red}$ is well approximated by quadratic fits, though $\chi^{2}_{red}$ from individual diagnostics can show very different behaviors. Furthermore, the fiducial configuration is close to the minima of the quadratic fits.

\begin{deluxetable}{ccccl}
  \tabletypesize{\textwidth}
  \tablecolumns{5}
  \tablewidth{0pt}
  \tablecaption{Minimum and 95\%-confidence limits of the quadratic fits (Fig.~\ref{fig:onlyChiLines}) and parameter values for the fiducial configuration} 
  \tablehead{\colhead{Parameter} & \colhead{Fiducial value} & \colhead{Central value} & \colhead{95\% CL} & \colhead{Description}}
  \startdata
    $mag_{0}$ & 30.565 & 30.576 & $\pm 0.055$ & Magnitude zero point \\
    $\sigma_{N}$ & 5.10 & 5.17 & $\pm 0.34$ & rms of the background noise \\
    $e_{1,rms}$ & 0.383 & 0.396 & $\pm 0.045$ & rms of the intrinsic $e_{1}$ distribution \\
    $e_{2,rms}$ & 0.407 & 0.387 & $\pm 0.041$ & rms of the intrinsic $e_{2}$ distribution \\
    $\sigma$ & 0.2422 & 0.2381 & $\pm 0.0390$ & rms of the intrinsic log-normal size distribution \\
    \multirow{2}{*}{$\theta$} & \multirow{2}{*}{0.1382} & \multirow{2}{*}{0.1370} & \multirow{2}{*}{$\pm 0.0068$} & \multirow{2}{*}{\begin{minipage}{2.7in}Rotation angle to plane intrinsic magnitudes and sizes are uncorrelated\end{minipage}}\\
  \enddata
  \label{tab:paramValues}
\end{deluxetable}

We want to point out a few features in the individual subfigures. First, the magnitude-size plane (see Fig. \ref{fig:diagMagSize}) and the ellipticity plane (see Fig. \ref{fig:diagEllip}) do not react to changes in the magnitude zero point of the image ($mag_{0}$). Only the histogram of pixel values (see Fig. \ref{fig:diagPixels}) responds to changes in $mag_{0}$, as the magnitude zero point affects the normalization of the pixel values. 

Second, when varying the width of the Gaussian background $\sigma_{N}$, the histogram of pixel values and the magnitude-size plane respond in opposite directions. As described in Section~\ref{sec:cl1}, the fiducial model produces a slightly deeper image compared to data. Thus, the magnitude-size plane is pushing for a higher noise level. The histogram of pixel values however constrains the background peak in the sky-subtracted image and cannot accommodate larger $\sigma_{N}$ values. Therefore, we believe that including an additional Non-Gaussian noise term can reconcile the tension.

Third, the diagnostics are rather flat if $e_{1,rms}$ and $e_{2,rms}$ are varied. As a result, the 95\%-confidence limits for this parameter are relatively wide. Another less noisy diagnostic to constrain the ellipticity distribution might shrink these confidence limits.

Fourth, the $\chi^{2}_{red}$ values of the ellipticity diagnostic are below 1 in the relevant parameter ranges. As the quotient in Eq.~\ref{eq:reducedChi} is dominated by the error estimated on the data $\sigma_{d,ij}$, this behavior of the ellipticity diagnostics suggests issues in estimating the scatter of the dataset.

Finally, the magnitude-size plane is the most sensitive diagnostic to changes in the magnitude and size distribution parameters $\sigma$ and $\theta$ of the galaxy population. Hence, the combined $\chi_{red}^{2}$-curve is mostly driven by this diagnostic.

\end{document}